\documentclass[11pt]{article}
\title{Exact uncertainty relations\\ - technical details}
\author{Michael J. W. Hall\\
Theoretical Physics, IAS\\Australian National University\\
Canberra ACT 0200, Australia}
\date{}
\begin{document}
\maketitle
%\newpage

\begin{abstract}
The Heisenberg inequality $\Delta X\Delta P\geq\hbar/2$ 
can be replaced by an exact equality, for suitably
chosen measures of position and momentum uncertainty, which is valid for
all wavefunctions.  The significance of this ``exact" uncertainty relation is
discussed, and results are generalised to angular momentum and phase, photon
number and phase, time and
frequency, and to states described by density operators.  Connections to
optimal estimation of an observable from the measurement of a second
observable, Wigner functions, energy 
bounds and entanglement are 
also given.  
\end{abstract}
\renewcommand{\thesection}{\Roman{section}}
\renewcommand{\thesubsection}{\Alph{subsection}}
\newpage
\section{INTRODUCTION}
One of the remarkable features of quantum mechanics is the property that
certain observables cannot simultaneously be assigned arbitrarily precise
values.  This property does not compromise claims of completeness for the
theory, since it may consistently be asserted that such observables cannot
simultaneously be {\it measured} to an arbitrary accuracy \cite{heisbohr}.
Thus the Heisenberg inequality
\begin{equation} \label{heis}
\Delta X \Delta P \geq \hbar /2
\end{equation}
is generally taken to 
reflect an essential incompleteness in the applicability of classical
concepts of position and momentum to physical reality.

It was recently noted that this fundamental inequality can 
be greatly strengthened.  In particular, one may
define a measure of position uncertainty $\delta X$ (which arises naturally in
classical statistical estimation theory), and a measure of nonclassical
momentum uncertainty $\Delta P_{nc}$ (which arises from a natural
decomposition of the momentum operator), such that \cite{hall} 
\begin{equation} \label{ex}
\delta X \Delta P_{nc} = \hbar /2 
\end{equation}
for {\it all} wavefunctions.
Such an equality may be regarded as an {\it exact} uncertainty relation, and
may be shown to imply the usual Heisenberg inequality Eq. (\ref{heis}).
Hence, perhaps paradoxically, the uncertainty principle
of quantum mechanics may be given a quantitatively precise form.

In Ref. \cite{hall} 
the above exact uncertainty relation was merely noted
in passing, with the emphasis being on
other properties of $\delta X$ and
$\Delta P_{nc}$. 
Similarly, while the very existence of
an exact form of the uncertainty principle was recently shown
to provide a sufficient basis for 
moving from the classical equations of motion to the Schr\"{o}dinger equation
\cite{hallreg}, the corresponding exact uncertainty relation Eq. (\ref{ex})
was only briefly mentioned.  The purpose of this paper, therefore, is to study
the physical significance of Eq. (\ref{ex}) in some detail, 
including its extensions to 
other pairs of conjugate observables and to general states described by
density operators.

In the following section it is shown that quantum observables such as
momentum, position, and photon number have a natural decomposition, into
the sum of a classical and a 
nonclassical component.  The classical component 
corresponds to the best possible measurement of the
observable, on a given state, which is compatible with measurement of
the conjugate observable.  Complementarity implies
that the classical component cannot be equivalent to
the observable itself, i.e., there is in general an nontrivial
{\it non}classical component.  It is this nonclassical component which
reflects the mutual incompatibility of pairs of conjugate observables,
and the magnitude of which appears in the exact uncertainty relations to
be derived [e.g., $\Delta P_{nc}$ in Eq. (\ref{ex})]. 
The decomposition into classical and nonclassical components is also related
in a natural manner to quantum continuity equations and to
quasiclassical properties of the Wigner function.

In Sec. III a measure of uncertainty is defined for continuous
random variables, which plays a fundamental role in classical estimation
theory, and which also provides a direct measure of the robustness of
the variable with respect to Gaussian diffusion processes.  This
measure, the ``Fisher length'' of the variable, may of course be
calculated for quantum observables as well, and appears as $\delta X$ in
the exact uncertainty relation in Eq. (\ref{ex}).

The ingredients of classical/nonclassical decompositions and
Fisher lengths are combined in Sec. IV to obtain a number of exact uncertainty
relations, such as Eq. (\ref{ex}) and the equality
\[ \delta \Phi \Delta N_{nc}=1/2\]
for phase and photon number, valid for all pure states.  These relations
generalise to {\it in}equalities for states described by density operators,
and are far stronger than the corresponding Heisenberg-type
inequalities.   It is shown that a bound on Fisher length leads to an
entropic lower bound for the groundstate energies of quantum systems,
and results are generalised to an exact uncertainty relation for time
and frequency, and to higher dimensions.

In Sec. V it is shown that the decomposition of an observable of
a given quantum system into
classical and nonclassical components is essentially nonlocal in nature,
being dependent in general on manipulations performed on a  
second system with which the first is entangled.  The significance of
the relevant exact uncertainty relations is discussed, with particular
reference to EPR-type states.

A formal generalisation of exact uncertainty relations, to
arbitrary pairs of quantum observables, is noted in Sec. VI.  Moreover,
it is shown that a result of Ivanovic \cite{ivan}, for complete sets of
mutually complementary observables on finite Hilbert spaces (such as the
Pauli spin matrices), may be reinterpreted as an exact uncertainty
relation for the ``collision lengths'' of the observables.

Conclusions are given in Sec. VII.

\section{CLASSICAL AND NONCLASSICAL COMPONENTS OF QUANTUM OBSERVABLES}

\subsection{Momentum}

The nonclassical momentum uncertainty 
$\Delta P_{nc}$ appearing in Eq. (\ref{ex}) is
defined via a natural decomposition of the momentum observable $P$ into
``classical'' and ``nonclassical'' components,
\begin{equation} \label{pdecomp}
P = P_{cl} + P_{nc} .
\end{equation}
This decomposition is state-dependent, and will be defined explicitly
further below.  In particular, it will be shown that {\it the classical
component, $P_{cl}$, corresponds to the best estimate of momentum for a given
quantum state compatible with a position measurement}. Moreover, 
the average error of this best estimate will be shown to correspond to
the variance $(\Delta P_{nc})^2$ of the nonclassical component.   In
Secs. II.B and II.C it will further be shown that $P_{cl}$ is related to
the momentum flow in a classical continuity equation following from the
Schr\"{o}dinger equation, and to an average momentum arising naturally
from quasiclassical properties of the Wigner function.  However, it is
the ``best estimate'' interpretation above that provides the most
general basis for generalisation to other observables. 

As a starting point, recall that in classical mechanics one can
simultaneously obtain precise values for position and momentum, whereas
in quantum mechanics one must choose to accurately measure either one or
the other.  It is therefore reasonable to ask the following question:
If I measure one of these observables precisely, on a known quantum
state, then what is the best
estimate I can make for the value of the other observable? Such an
estimate of momentum from the measurement of position will be called a
{\it classical} estimate of $P$, since it assigns simultaneous values to
$X$ and $P$.

It will be shown that the {\it best} classical estimate of $P$, given
the measurement result $X=x$ on a quantum system described by wavefunction
$\psi (x)$, is given by
\begin{equation}
\label{pclpsi}
P_{cl}(x) = \frac{\hbar}{2i}
\left( \frac{\psi^\prime (x)}{\psi(x)}-\frac{\psi^{*\prime} (x)}{\psi^*(x)}
\right) = \hbar [\arg \psi (x)]^\prime .
\end{equation}
More generally, for a quantum system described by density operator
$\rho$, one has
\begin{equation} \label{pcl}
P_{cl}(x) := \frac{\langle x|P\rho +\rho P|x\rangle/2}{\langle x|\rho
|x\rangle} 
\end{equation}
(which reduces to the first expression for $\rho =|\psi\rangle\langle\psi
|$).  Note that this estimate is equivalent to measurement of the
Hermitian operator 
\begin{equation} \label{pclop}
P_{cl} = \int dx\,  P_{cl}(x)|x\rangle\langle x| 
\end{equation}
on state $\rho$, which by construction commutes with $X$.  
The experimentalist's procedure is thus to (i) prepare the system in
state $\rho$; (ii) measure the position $X$; and (iii) for result $X=x$
calculate $P_{cl}(x)$.  As stated above, this procedure yields the best
possible estimate of the momentum of the system that is compatible with
simultaneous knowledge of the position of the system.

It is important to note that $P_{cl}(x)$ and $P_{cl}$ should, strictly
speaking, explicitly indicate their dependence on a given state $\rho$,
e.g., via the notation $P_{cl}(x|\rho)$ and $P_{cl}^\rho$ respectively.
This would in particular be necessary if one wished to evaluate the
expectation value ${\rm tr}[\sigma P_{cl}^\rho]$ for some density operator
$\sigma$ other than $\rho$.  
However, since in fact expectation values will only be
evaluated for the corresponding state $\rho$ throughout this paper,
explicit notational dependence on the state may be conveniently
dispensed with, without
leading to ambiguity.  Similar remarks
apply to the nonclassical momentum component $P_{nc}$ in Eq.
(\ref{pdecomp}).

To prove that $P_{cl}(x)$ above provides the best classical estimate of
$P$, consider some general classical estimate
$\tilde{P}(x)$ for momentum associated with measurement result $X=x$ for
state $\rho$.  This estimate is then equivalent to measurement of the
operator $\tilde{P} = \int dx\,\tilde{P}(x)|x\rangle\langle x|$,
and hence the average error of the estimate may be quantified by
\begin{equation} \label{error1}
{\cal E}_P := \langle (P-\tilde{P})^2\rangle = \langle P^2\rangle +
\langle \tilde{P}^2\rangle - \langle \tilde{P}P+P\tilde{P}\rangle ,
\end{equation} 
where $\langle A\rangle$ denotes ${\rm tr}[\rho A]$.  
But, using the cyclic property of the trace operation
and evaluating the trace in the position representation,
\begin{eqnarray*}
\langle \tilde{P}P+P\tilde{P}\rangle & = & \int dx\,\langle x|\tilde{P}
P\rho + \rho P\tilde{P}|x\rangle\\
& = & \int dx\, \tilde{P}(x)\langle x|P\rho +\rho P|x\rangle\\
& = & 2\int dx\, \langle x|\rho |x\rangle \tilde{P}(x) P_{cl}(x) =
2\langle\tilde{P}P_{cl}\rangle ,
\end{eqnarray*}
and hence
\begin{eqnarray}
{\cal E}_P & = & \langle P^2\rangle +\langle \tilde{P}^2\rangle -
2\langle\tilde{P}P_{cl}\rangle\nonumber \\ \label{error2}
& = & \langle P^2\rangle - \langle P_{cl}^2\rangle +\langle (\tilde{P}
- P_{cl})^2\rangle \label{error} .
\end{eqnarray}
Since the last term is positive, the average error is therefore
minimised by the choice $\tilde{P}=P_{cl}$ as claimed.

The nonclassical momentum component $P_{nc}$ is implicitly defined via
Eqs. (\ref{pdecomp}) and (\ref{pclop}).  
From Eq.~(\ref{pcl}) one finds that the expectation values of
the observables $P$ and $P_{cl}$ are always equal (for the corresponding
state $\rho$), i.e.,
\begin{equation} \label{pav}
\langle P\rangle = \langle P_{cl}\rangle ,\hspace{1cm}\langle
P_{nc}\rangle = 0 . 
\end{equation}
Hence the quantum momentum $P$ in Eq. (\ref{pdecomp}) 
can also be interpreted as the sum of
an average momentum, $P_{cl}$, 
and a nonclassical momentum fluctuation, $P_{nc}$.
Moreover, the magnitude of this fluctuation is simply related to the
minimum average error:  choosing $\tilde{P}=P_{cl}$ 
implies from Eqs. (\ref{pdecomp}), (\ref{error1}) and (\ref{error2}) that
\begin{equation} \label{pdiff}
{\cal E}_P^{\rm min}
= \langle(P-P_{cl})^2\rangle = \langle P_{nc}^2
\rangle = \langle P^2\rangle - \langle P_{cl}^2\rangle .
\end{equation} 
It will be seen that, as a consequence of the exact uncertainty relation
Eq. (\ref{ex}), this error does not vanish for any state (although it
may be arbitrarily small).
Note from Eqs. (\ref{pav}) and (\ref{pdiff}) that the nonclassical
fluctuation strength $\Delta P_{nc}$ in Eq. (\ref{ex}) is a fully operational
quantity, as it may be determined from the measured distributions of $P$
and $P_{cl}$. 

Several formal properties further support the physical significance
of the decomposition in Eq. (\ref{pdecomp}).  First, the classical and
nonclassical components are linearly uncorrelated, i.e.,
\begin{equation} \label{pvar}
{\rm Var} P = {\rm Var} P_{cl} + {\rm Var} P_{nc} , 
\end{equation}
as follows immediately from Eqs. (\ref{pav}) and
(\ref{pdiff}).  This
implies a degree of statistical, and hence physical,
independence for $P_{cl}$ and $P_{nc}$.  Second, the classical momentum
component commutes with the conjugate
observable $X$ while the nonclassical component does not, i.e.,
\begin{equation}
[X,P_{cl}]=0,\hspace{1cm} [X,P_{nc}] = i\hbar .\nonumber 
\end{equation}
Hence it is the {\it nonclassical} component of $P$ which generates
the fundamental quantum property $[X,P]=i\hbar$. 
Finally, when the decomposition is
generalised to more than one dimension (see Sec. IV.E), one finds that
the commutativity property $[P^j,P^k]=0$ for the vector components of
momentum is preserved by the decomposition, i.e.,  
\begin{equation}\label{commcomp}
[P_{cl}^j, P_{cl}^k] = 0 = [P_{nc}^j, P_{nc}^k].\nonumber
\end{equation}

The decomposition in Eq. (\ref{pdecomp}) attempts to demarcate
classical and nonclassical
momentum properties. 
It is therefore reasonable to hope that the {\it nonclassical}
component $P_{nc}$ in particular might play a fundamental role in describing the
essence of what is ``quantum'' about quantum mechanics.  This is indeed
the case.  A derivation of the Schr\"{o}dinger equation
as a consequence of adding a nonclassical momentum fluctuation to a classical
ensemble (with strength inversely proportional to the uncertainty in position),
has recently been given
\cite{hallreg}.  In this paper it will be shown that the nonclassical 
components of quantum observables, such as position, momentum
and angular momentum,  
satisfy {\it exact} uncertainty relations
such as Eq. (\ref{ex}).  It will further be shown that the
decomposition of observables into classical and nonclassical components
helps to distinguish between local and nonlocal features of quantum
entanglement.

\subsection{Angular momentum}

Angular momentum takes quantized values in quantum mechanics, but
continuous values in classical mechanics.  Hence it is not immediately
clear whether a decomposition into classical and nonclassical
contributions can exist, analogous to Eq. (\ref{pdecomp}).  A similar
remark may be made for photon number. However, it will be seen that
discreteness {\it per se } imposes no impediment (see also Sec. VI).  

For simplicity, consider a rigid rotator confined to the $xy$-plane,
with angular momentum
\[
J=J_z=\frac{\hbar}{i}\frac{\partial}{\partial\phi},
\]
moment of inertia $I$, and phase angle $\phi$.  If a phase-dependent
potential $V(\phi)$ acts on the rotator (eg, $V(\phi)=mg\cos \phi$ for a
pendulum), then the corresponding Hamiltonian is 
\[
H= J^2/(2I) + V(\phi) .
\]

A pure state of the rotator has corresponding angular momentum and phase
representations
\[
|\psi\rangle = \sum_j \psi_j |j\rangle =
\int_0^{2\pi} d\phi \, f(\phi)|\phi\rangle ,
\]
where $|j\rangle$ is the eigenstate of angular momentum $\hbar j$,
$|\phi\rangle$ is the phase eigenket $(2\pi)^{-1/2}\sum_j
e^{-ij\phi}|j\rangle$, and the phase wavefunction $f(\phi)$ is related to the
amplitudes $\psi_j$ by
\[
f(\phi)= \langle\phi |\psi\rangle = (2\pi)^{-1/2} \sum_j \psi_j e^{ij\phi}  .
\]

By analogy with Eq. (\ref{pdecomp}), the angular momentum can be
decomposed into classical and nonclassical components,
\begin{equation} \label{jdecomp}
J = J_{cl} + J_{nc} , 
\end{equation}
with
\begin{eqnarray} \label{jclop}
J_{cl} & = & \int d\phi\, J_{cl}(\phi) |\phi\rangle\langle \phi |\\
J_{cl}(\phi) & = & \langle\phi |J_{cl} |\phi\rangle = \frac{\langle
\phi |J\rho +\rho
J|\phi\rangle/2}{\langle \phi |\rho |\phi\rangle}  \label{jcl}
\end{eqnarray}
in analogy to Eqs. (\ref{pclop}) and (\ref{pcl}) respectively.
One may show that $J_{cl}(\phi)$ is the best estimate of angular
momentum compatible with a measurement of phase for state $\rho$, and that
\begin{eqnarray}
\langle J\rangle & = & \langle J_{cl}\rangle ,\hspace{1cm}
\langle J_{nc}\rangle = 0 \label{jav},\\
{\rm Var} J & = & {\rm Var} J_{cl} + {\rm Var} 
J_{nc} \label{jvar}
\end{eqnarray}
in analogy to Eqs. (\ref{pav}) and (\ref{pvar}).
Note from these properties that one also has
\[ \langle J^2\rangle = \langle J_{cl}^2\rangle +
\langle J_{nc}^2\rangle , \]
and hence the kinetic energy $\langle J^2\rangle /(2I)$ splits into a
classical contribution and a nonclassical contribution.
An exact uncertainty relation for $J_{nc}$ and phase angle will be
derived in
Sec. IV.

It is of interest to point out an alternative approach to the
decomposition in Eq. (\ref{jdecomp}), based
on the continuity equation for the phase probability density.
In particular, restricting to a pure state $\rho=|\psi\rangle\langle\psi
|$ for convenience,
multiplying the Schr\"{o}dinger equation for the phase wavefunction $f(\phi)$
by $f^*(\phi)$ and taking the imaginary part 
yields the continuity equation
\begin{equation} \label{phasecont}
\partial |f|^2/\partial t + (\partial/\partial\phi)[|f|^2 I^{-1}
J_{cl}(\phi)] = 0 ,
\end{equation}
with $J_{cl}(\phi)$ defined as above.
Thus $I^{-1}J_{cl}$ 
is the angular stream velocity associated with members of 
a classical ensemble of rotators
described by phase density $|f|^2$, and hence 
$J_{cl}$ is the corresponding angular
momentum.  

A similar ``dynamical'' approach, based on the continuity equation \cite{merz}
\[
\partial |\psi |^2/\partial t + (\partial/\partial x)[|\psi |^2 m^{-1}
P_{cl}(x)] = 0  \]
for the position probability density,
was given in Ref. \cite{hall} as the basis for defining the
momentum decomposition of Eq. (\ref{pdecomp}).  
However, such approaches are in general only applicable for
systems with Hamiltonians quadratic in the observable of interest.

\subsection{Wigner function approach}

In this subsection another approach to the decomposition of
position and momentum observables is noted, 
based on an analogy between
classical phase space distributions and the Wigner function.
In this approach $P_{cl}$ appears as the natural 
quantum analogue of a classical average momentum.  

The Wigner function $W(x,p)$ corresponding to density operator $\rho$
is defined by \cite{wig}
\begin{equation} \label{wig}
W(x,p) := (2\pi\hbar)^{-1}\int d\xi 
e^{-ip\xi/\hbar}\langle x-\xi/2|\rho |x+\xi/2\rangle ,
\end{equation}
and behaves like a joint probability density for position and momentum
to the extent that 
\begin{eqnarray*}
\langle x|\rho |x\rangle & = & \int dp \, W(x,p) \\
\langle p|\rho |p\rangle & = & \int dx \, W(x,p) .
\end{eqnarray*}
However, $W(x,p)$ can typically take negative values, and is
hence fundamentally nonclassical in nature.

Now, if $\rho(x,p)$ is a classical joint probability density on phase
space, then one can define the {\it average} momentum associated with position
$x$ by 
$p_{av}(x)=\int dp\, p\, {\rm prob}(p|x)$, 
where ${\rm prob}(p|x)$ denotes
the conditional probability that the momentum is equal to $p$ at
position $x$, i.e.,
${\rm prob}(p|x) = \rho(x,p)/\int dp\,\rho(x,p)$. 
The classical average momentum at position $x$ is thus 
\[
p_{av}(x) = \frac{\int dp\, p\rho (x,p)}{\int dp \, \rho (x,p)}.
\]

This immediately suggests defining an analogous {\it quantum} average momentum
associated with position $x$ by \cite{takbrown}
\begin{equation} \label{qav}
P_{av}(x) := \frac{\int dp\, p W(x,p)}{\int dp \, W(x,p)} ,
\end{equation}
yielding a natural decomposition of the momentum observable $P$
into an average component and a fluctuation component:
\begin{equation} \label{avdecomp}
P = P_{av} + P_{fluc} , 
\end{equation} 
where $P_{av}=\int dx\, P_{av}(x)|x\rangle\langle x|$.

Remarkably, this is equivalent to the decomposition in Eq.
(\ref{pdecomp}).  In particular, one has the identities
\begin{equation} \label{clav}
P_{av} \equiv P_{cl},\hspace{1cm} P_{fluc}\equiv P_{nc} . 
\end{equation}
This follows by first substituting Eq. (\ref{wig}) into Eq. (\ref{qav}) 
and using integration by parts, to give
\begin{eqnarray*}
\langle x|\rho |x\rangle P_{av}(x) & = & (2\pi\hbar)^{-1} \int\!\!\int dp d\xi \left[
i\hbar \frac{d}{d\xi} e^{-ip\xi/\hbar}\right]\langle x-\xi/2|\rho|x+\xi/2
\rangle\\
& = & (\hbar/i) \int d\xi\, \delta(\xi)
(d/d\xi)\langle x-\xi/2|\rho|x+\xi/2\rangle\\
& = & \left. (\hbar/i)\frac{d}{d\xi}\langle x-\xi/2|\rho
|x+\xi/2\rangle\right|_{\xi=0} . 
\end{eqnarray*}
Expanding in momentum eigenkets then yields
\begin{eqnarray*}
\langle x|\rho |x\rangle P_{av}(x) & = & 2^{-1}\int\!\!\int dpdp'\,
(p+p')
\langle p|\rho |p'\rangle e^{ix(p-p')/\hbar}\\
& = & 2^{-1}\int\!\!\int dpdp'\, \langle p|P\rho +\rho P|p'\rangle
e^{ix(p-p')/\hbar}\\
& = & \langle x|P\rho + \rho P|x\rangle/2 
\end{eqnarray*}
as required.

The Wigner function thus enables an alternative approach to the
decomposition in Eq. (\ref{pdecomp}), which moreover reinforces the
interpretation of Eq. (\ref{pav}), that the momentum of a
quantum particle comprises a nonclassical fluctuation about a classical
average. 
As an immediate application, note that in obvious analogy to Eqs.
(\ref{qav})-(\ref{clav}) one may define the corresponding decomposition
of the position observable $X$ into classical and nonclassical
components via
\begin{eqnarray} \label{xdecomp}
X & = & X_{cl} + X_{nc},\\
X_{cl}(p) & = & \frac{\int dx\, x W(x,p)}{\int dx\, W(x,p)} , 
\nonumber\end{eqnarray}
where $X_{cl}=\int dp\, X_{cl}(p)|p\rangle\langle p|$.
This agrees with the analogous definition based on Eq. (\ref{pcl}),
corresponding to a ``best estimate'' approach, and also 
with the definition given 
in Ref. \cite{hall} based on a semiclassical continuity equation.

\subsection{Photon number}

Determining a classical component of the photon number $N$ is reasonably
straightforward.  However, because the
observable conjugate to $N$ is not represented by a Hermitian operator,
the notion of a decomposition $N=N_{cl}+N_{nc}$ has to be generalised.
The reader not interested in the technical details of this generalisation
may wish merely to note Eqs. (\ref{ncl}), (\ref{nav}) and (\ref{nvar})
below, which are analogous to Eqs. (\ref{jcl}), (\ref{jav}) and
(\ref{jvar}) respectively.

The most general description of an observable $A$, consistent with
standard quantum theory, is via a {\it probability operator measure}
(POM), i.e., via a set of positive operators $\{ A_j\}$ which sum to the
identity operator \cite{helhol}.  The probability of result
$A=a_j$ for a measurement of $A$ on state $\rho$ is given by ${\rm
tr}[\rho A_j]$.  For the special case of an observable with a Hermitian
operator representation, $A_j$ is just the projection onto the eigenspace
corresponding to $a_j$.

The phase observable $\Phi$ conjugate to the photon number observable
$N$ is described by the continuous POM $\{ |\phi\rangle\langle\phi |\}$
with \cite{helhol,shap,hallqo}
\begin{equation} \label{nphi}
|\phi\rangle := (2\pi)^{-1/2} \sum_n e^{in\phi}|n\rangle ,
\end{equation}
where $|n\rangle$ denotes the eigenstate corresponding to $n$ photons.
The probability density for obtaining phase value $\phi$ for a
measurement of $\Phi$ on state $\rho$ is therefore
\begin{equation} \label{phidist}
p(\phi |\rho) = {\rm tr}[\rho |\phi\rangle\langle\phi |] = \langle\phi
|\rho |\phi\rangle .
\end{equation}
The phase kets $|\phi\rangle$ may be recognised as eigenkets of
the (non-Hermitian) Susskind-Glogower phase operator \cite{susslevy},
and are not mutually orthogonal.

As per Sec. II.A, one may consider a classical estimate
$\tilde{N}(\phi)$ of photon number based on measurement result $\Phi
=\phi$ for state $\rho$.  Note that such an estimate corresponds to a
POM observable $\tilde{N}$, with measurement outcome determined by a
measurement of $\Phi$.  Thus $\tilde{N}$ and $\Phi$ are compatible
observables, being jointly measurable.

To determine the {\it best} classical estimate of $N$, one has to choose
an appropriate measure of error.  Here a difficulty arises:  one
cannot in general add or subtract POM observables as they do not have
algebraic representations as operators.  Hence the expression $\langle
(N-\tilde{N})^2\rangle$ analogous to Eq. (\ref{error1}) is not well
defined.  However, evaluating Eq. (\ref{error1}) in the position
representation yields the equivalent expression
${\cal E}_P = \int dx\, \langle x|[P-\tilde{P}(x)]
\rho[P-\tilde{P}(x)]|x\rangle$, 
and hence one may analogously define
\begin{equation} \label{nerror1}
{\cal E}_N = \int d\phi\, \langle\phi |[N-\tilde{N}(\phi)]
\rho[N-\tilde{N}(\phi)]|\phi
\rangle
\end{equation} 
for the average error of a classical estimate of $N$.  It follows,
precisely as per the minimisation of ${\cal E}_P$ in Sec. II.A, that the
{\it best} classical estimate of photon number is given by
\begin{equation} \label{ncl}
N_{cl}(\phi) := \frac{\langle\phi |N\rho +\rho N|\phi\rangle
/2}{\langle\phi |\rho |\phi\rangle}  .
\end{equation}

The classical photon number observable, $N_{cl}$, thus shares formal
similarities with $P_{cl}$ and $J_{cl}$ in Eqs. (\ref{pcl}) and
(\ref{jcl}) respectively.  Moreover, it is straighforward to show that
\begin{equation} \label{nav}
\langle N\rangle = \langle N_{cl}\rangle
\end{equation}
in analogy to Eqs. (\ref{pav}) and (\ref{jav}).  However, the algebraic
difficulty mentioned above again arises in the definition of a
corresponding {\it nonclassical} photon number observable $N_{nc}$.  In
particular, the formal expression
\[
N = N_{cl} + N_{nc} \]
is not well defined, since $N_{cl}$ does not have a Hermitian operator
representation in general.  This difficulty does not in fact pose a
problem for obtaining an exact uncertainty relation for phase and photon
number (as ${\rm Var}N-{\rm Var}N_{cl}$ can be substituted for
$(\Delta N_{nc})^2$), but for completeness will be
resolved further below.

Note first that, irrespective of the existence  of a formal
decomposition into classical and nonclassical observables, one can
define a decomposition of the average energy $\langle H\rangle =
\hbar\omega\langle N+1/2\rangle$ into classical and quantum components
by
\begin{equation} \label{edecomp}
\langle H\rangle = E_{cl} + E_{nc} ,
\end{equation}
where $E_{cl}:=\hbar\omega\langle N_{cl}\rangle$.  For the particular
case of an eigenstate of $n$ photons it follows via Eq. (\ref{ncl}) that
\[
E_{cl} = n\hbar\omega , \hspace{1cm} E_{nc} = \frac{1}{2}\hbar\omega .\]
Thus the nonclassical energy is precisely the vacuum energy for such
states.

Finally, to define a POM observable $N_{nc}$ which can be regarded as
representing
the nonclassical component of photon number, it is simplest to exploit
formal similarities between photon number and angular momentum.
In particular, extend the Hilbert space to include a set of ``negative
photon number'' states $\{|n\rangle :\, n=-1,-2,-3,\dots\}$, and define
the extended photon operator \cite{pegg}
\[
N^* = \sum_{n=-\infty}^\infty n|n\rangle\langle n| \]
and the (mutually orthogonal) extended phase states
\[
|\phi^*\rangle = (2\pi)^{-1/2}\sum_{n=-\infty}^\infty
e^{in\phi}|n\rangle .  \]
This is formally analogous to the case of angular momentum considered in
Sec. II.B, and in particular one may define the operator decomposition
\[
N^* = N^*_{cl} + N^*_{nc} \]
analogous to Eq. (\ref{jdecomp}). 

Consider now the projection operator 
$E=\sum_{n=0}^\infty |n\rangle\langle n|$, 
which projects onto the original Hilbert space. 
For any ``physical'' state, i.e., any
state with no negative photon number components, one has $\rho=E\rho E$.
Thus, substituting $E\rho E$ for $\rho$ in the expression for
$N^*_{cl}(\phi)$ analogous to Eq. (\ref{ncl}), and noting the identities
$E|\phi^*\rangle=|\phi\rangle$, $EN^*=N=NE$ and their conjugates, one
finds that $N^*_{cl}(\phi)=N_{cl}(\phi)$ for such states.
Moreover, 
any given POM observable $\{A^*_j\}$ on the extended Hilbert
space is mapped by $E$ 
to the POM observable $\{A_j\equiv EA^*_jE\}$ on the
original Hilbert space. 
It is straightforward to check that $E$ maps $N^*$ and
$N^*_{cl}$ to $N$ and $N_{cl}$ respectively (for states
with no negative energy components).  Hence one may {\it define}
the observable $N_{nc}$ as the POM mapped to by $E$ from
$N^*_{nc}$ (i.e., as the POM obtained by applying $E$ to the
projections onto the eigenspaces of $N^*_{nc}$).
 
Under the above definition of $N_{nc}$
one has the statistical independence property
\begin{equation} \label{nvar}
{\rm Var} N = {\rm Var} N_{cl} + {\rm Var} N_{nc}
\end{equation}
in analogy to Eqs. (\ref{pvar}) and (\ref{jvar}).  Indeed, this result
is a trivial consequence of the corresponding relation for $N^*$,
$N^*_{cl}$ and $N^*_{nc}$.  More generally, for any state $\rho$ with no
negative photon number components one has
\[
{\rm tr}[\rho A^*_j] = {\rm tr}[E\rho EA^*_j] = {\rm tr}[\rho EA^*_jE] =
{\rm tr}[\rho A_j] \]
for any POM observable $A^*$.  The statistical properties of $A^*$ and
$A$ are therefore {\it identical} for any such state.  Thus $N_{cl}$
and $N_{nc}$ inherit all statistical properties of $N^*_{cl}$ and
$N^*_{nc}$ respectively, including Eq. (\ref{nvar}).

\section{FISHER LENGTH}

\subsection{Position}

The uncertainty measure $\Delta P_{nc}$ in Eq. (\ref{ex}) is now well
defined - it is the rms uncertainty of the nonclassical momentum component
$P_{nc}$.  However, it still remains to define the measure of position
uncertainty $\delta X$ in Eq. (\ref{ex}).  This is done below for the
general case of continuous observables taking values over the entire set
of real numbers, such as position and momentum, while the case of
periodic observables such as phase is treated in Sec. III.B.  Note
that $\delta X$ is a purely {\it
classical} measure of uncertainty, requiring no reference to quantum
theory whatsoever.  

For a random variable $X$ which takes values over the whole range of
real numbers, there are of course many possible ways to quantify the
spread of the corresponding distribution $p(x)$.  Thus, for example, one
may choose the rms uncertainty $\Delta X$, the collision length $1/\int
dx \, p(x)^2$ \cite{hell}, or the ensemble length $\exp [-\int dx\, p(x) \ln
p(x)]$ \cite{hallvol}.  All of these examples have the desirable
properties of having the same units as $X$, scaling with $X$, and
vanishing in the limit as $p(x)$ approaches a delta function.

A further uncertainty measure satisfying the above
properties is
\begin{equation} \label{fl}
\delta X := \left[ \int_{-\infty}^{\infty} dx \, p(x) \left( \frac{d\ln
p(x)}{dx}\right)^2\right]^{-1/2}  .
\end{equation}
While this measure may appear unfamiliar to physicists, it is in fact
closely related to the well known Cramer-Rao inequality that lies at
the heart of statistical estimation theory \cite{cox}: 
\begin{equation} \label{cramer}
\Delta X \geq \delta X . 
\end{equation}
Thus $\delta X$ provides a lower bound for $\Delta X$. 
Indeed, more generally, $\delta X$ provides 
the fundamental lower bound for the rms
uncertainty of {\it any} unbiased estimator for $X$ \cite{cox}.  The bound 
in Eq. (\ref{cramer}) is
tight, being saturated if and only if $p(x)$ is a Gaussian distribution.

Eq. (\ref{cramer}) is more usually written in the form ${\rm Var} X\geq1/F_X$,
where $F_X=(\delta X)^{-2}$ is the ``Fisher information'' associated
with translations of 
$X$ \cite{cox,fish,stam}.  It is hence appropriate to refer to $\delta
X$ as the {\it Fisher length}.  
From Eq. (\ref{fl}) it is seen that the Fisher length may be regarded
as a measure of the length scale over which $p(x)$ (or, more precisely,
$\ln p(x)$) varies rapidly.

Basic properties of the Fisher length are: (i) $\delta Y=\lambda\delta X$ for
$Y=\lambda X$; (ii) $\delta X\rightarrow 0$ as $p(x)$ approaches a delta
function; (iii) $\delta X \leq \Delta X$ with equality only for Gaussian
distributions; and (iv) $\delta X$ is finite for all distributions.  
This last property follows since the integral in Eq. (\ref{fl}) can
vanish only if $p(x)$ is constant everywhere, which is inconsistent with
$\int dx\, p(x)=1$.

The Fisher length has the unusual feature that it depends on the
derivative of the distribution.   Moreover, for this reason it vanishes for
distributions which are discontinuous - to be expected from the above 
interpretation of $\delta X$, since such distributions vary
{\it infinitely} rapidly over a {\it zero} 
length scale ($\delta X=0$ may be shown by replacing such
a discontinuity at point $x_0$ by a linear interpolation over an
interval $[x_0-\epsilon, x_0+\epsilon]$ and taking the limit
$\epsilon\rightarrow 0$).  The Fisher length also vanishes
for a distribution that is zero over some interval (since $\ln p(x)$ in
Eq. (\ref{fl}) changes from $-\infty$ to a finite value over any
neighbourhood containing an endpoint of the interval).  
While these features imply that
$\delta X$ is not a particularly useful uncertainty measure for such
distributions (similarly, $\Delta X$ is not a particularly useful measure
for the Cauchy-Lorentz distribution $(a/\pi)(a^2+x^2)^{-1}$), they
are {\it precisely} the features that lead to a simple proof that the
momentum uncertainty is infinite for any quantum system with a position
distribution that is discontinuous or vanishes over some interval (as will
be shown in Sec. IV).

One further property of Fisher length worthy of note is its alternative
interpretation as a ``robustness length''.  In particular, suppose that
a variable described by $p(x)$  is subjected to a Gaussian diffusion
process, i.e., $\dot{p}=\gamma p''+\sigma p'$ for diffusion constant $\gamma$ 
and drift velocity $\sigma$.  
It then follows from Eq. (\ref{fl}) and de Bruijn's
identity \cite{stam} that the rate of entropy increase is given by
\begin{equation} 
\dot{S} = \gamma /(\delta X)^2 .
\end{equation}
Since a high rate of entropy increase corresponds to a rapid spreading of the
distribution, and hence nonrobustness to diffusion, this inverse-square law
implies that the Fisher length $\delta X$ is a direct measure of
robustness.  Hence $\delta X$ may also be referred to as a
{\it robustness length}.  This characterisation of robustness is 
explored for quantum systems in Ref. \cite{hall}.

Finally, note that Fisher length is not restricted to position
observables, but may be calculated as per Eq. (\ref{fl}) for any
observable which takes values over the entire set of real numbers, such
as momentum.

\subsection{Phase}

For a periodic random variable the corresponding Fisher
length is defined in a slightly modified manner, and satisfies a
correspondingly modified
Cramer-Rao inequality.  In particular, for a phase
variable $\Phi$ with associated period $2\pi$ and periodic phase
distribution $p(\phi)$ one defines
\begin{equation} \label{fphi}
\delta \Phi := \left[ \int_0^{2\pi} d\phi\, p(\phi)\left( \frac{d\ln
p(\phi)}{d\phi}\right)^2\right]^{-1/2} .
\end{equation}
This quantity satisfies many of the same properties as $\delta X$ above,
and again may be interpreted as a robustness length.
However, $\delta\Phi$ is distinguished from $\delta X$ in two important
respects. 

First, due to the compact support of $p(\phi)$, it is possible for
$p(\phi)$ to be a uniform distribution, with $\delta\Phi=\infty$. 
Thus $\delta\Phi$ perhaps somewhat overestimates the spread of a
uniform distribution ! (just as $\Delta X$ overestimates the spread of a
Cauchy-Lorentz distribution). Note this property implies that a uniform
phase distribution is infinitely robust to diffusion - it simply
cannot spread any further. This property is also  precisely what
is needed for the existence of an exact uncertainty relation between
phase and photon number, as will be seen in Sec. IV.

Second, and more importantly, $\delta\Phi$ satisfies a modified form of
the Cramer-Rao inequality in Eq. (\ref{cramer}).  In particular, for a
periodic phase distribution $p(\phi)$, define the ``variance''
about an arbitrary angle $\theta$ by \cite{hallphase}
\begin{equation} \label{varphi}
{\rm Var}_\theta \Phi := \int_{\theta-\pi}^{\theta+\pi} d\phi \,
(\phi-\theta)^2p(\phi) ,
\end{equation}
with corresponding rms uncertainty $\Delta_\theta \Phi$ \cite{footphase}.
One may then derive the Cramer-Rao type inequality
\begin{equation} \label{phicramer}
\Delta_\theta \Phi \geq |1-2\pi p(\theta +\pi)|\delta\Phi .
\end{equation}
Note that for a distribution highly peaked about a mean value $\theta$
one will typically have $p(\theta+\pi) << 1$, and hence this inequality
reduces to $\Delta_\theta\Phi\geq\delta\Phi$ in analogy to Eq.
(\ref{cramer}).

To obtain Eq. (\ref{phicramer}), note that integration by parts and the
periodicity of $p(\phi)$ gives 
\[
\int_{\theta-\pi}^{\theta+\pi} d\phi\, p'(\phi)(\phi-\theta) = 2\pi
p(\theta +\pi)-1 .
\]
But from the Schwarz inequality one has 
\begin{eqnarray*}
\left[ \int_{\theta-\pi}^{\theta+\pi} d\phi
\, p'(\phi)(\phi-\theta)\right]^2 & = & \left\{
\int_{\theta-\pi}^{\theta+\pi} d\phi\, \left[
p'(\phi)/\sqrt{p(\phi)}\right]\, \left[\sqrt{p(\phi)}(\phi-\theta)\right]
\right\}^2 \\
& \leq & \int_{\theta-\pi}^{\theta+\pi}d\phi \, p'(\phi)^2/p(\phi) 
\int_{\theta-\pi}^{\theta+\pi}d\phi \, p(\phi)(\phi-\theta)^2 . 
\end{eqnarray*}
Eq. (\ref{phicramer}) then follows via the definitions in Eqs.
(\ref{fphi}) and (\ref{varphi}).  Note that equality holds only in the
case that the Schwarz inequality is saturated, i.e., when the two
terms in square brackets in the first equality above are proportional.
This occurs when $p(\phi)$ is a (truncated) Gaussian or inverted Gaussian,
centred on $\theta$.

\section{EXACT UNCERTAINTY RELATIONS}

\subsection{Position and momentum}

In the previous two sections the quantities $\Delta P_{nc}$ and $\delta X$
have been motivated and discussed on completely independent grounds.
One is a measure of uncertainty for the nonclassical component
of momentum, while the other is a measure of uncertainty for position that
appears naturally in the contexts of classical statistical estimation
theory and Gaussian diffusion processes.

It is a remarkable fact that for all pure states these two quantities are
related by the simple equality
in Eq. (\ref{ex}), repeated here for convenience:
\begin{equation} \label{rex}
\delta X \Delta P_{nc} = \hbar/2 .
\end{equation}
Thus the Fisher length of position is inversely proportional to the
strength of the nonclassical momentum fluctuation.  
Note from Eqs. (\ref{pvar})
and (\ref{cramer}) that $\Delta P\geq\Delta P_{nc}$ and $\Delta X\geq
\delta X$ respectively.  Hence the Heisenberg uncertainty relation
Eq. (\ref{heis}) is an immediate consequence of this {\it exact}
quantum uncertainty relation.

A simple proof of Eq. (\ref{rex}) was given in Ref. \cite{hall}; a more
general result, valid for density operators, is proved below.
Before proceeding to the proof, however, several simple consequences 
of the exact uncertainty
relation in Eq. (\ref{rex}) are noted. 

First, recalling that $\delta X$ vanishes for position distributions
that are discontinuous or are zero over some interval (see Sec. III.A),
it follows immediately from Eq. (\ref{rex}) that $\Delta P_{nc}$ is
infinite in such cases.  From Eq. (\ref{pvar}) $\Delta
P$ is then also infinite.  Note that this conclusion {\it
cannot} be derived from the Heisenberg inequality Eq. (\ref{heis}), nor
from the entropic uncertainty relation for position and momentum
\cite{bbm}. The exact uncertainty relation Eq. (\ref{rex}) is thus
significantly stronger than the latter inequalities.  

A second related consequence worth mentioning is a simple proof that 
any well-localized state, i.e., one for
which the position distribution vanishes outside some finite interval,
has an infinite energy (at least for any potential energy that is bounded
below at infinity). This is immediately implied by the property 
\begin{equation}\label{energy}
E = (8m)^{-1}\hbar^2(\delta X)^{-2}
+ \langle P_{cl}^2\rangle /(2m) + \langle V(x)\rangle
\end{equation}
(following from Eqs. (\ref{pdiff}) and (\ref{rex})), 
noting that $\delta X=0$ for such states.
Note that this ``paradox'' of standard quantum mechanics 
(that there are no
states which are both well-localised and have finite energy) is
a consequence of the simple external potential model, rather than of some deep
incompleteness of the theory.  Note also that this property is purely
quantum in nature, since the divergent term vanishes in the limit
$\hbar\rightarrow 0$.

Third, the property $\delta X<\infty$ (see Sec.
III.A) immediately implies from the exact uncertainty relation Eq.
(\ref{rex}) that $\Delta P_{nc}$ can never vanish, i.e.,
\begin{equation}\label{posit}
\Delta P_{nc} > 0 .
\end{equation}
Thus all quantum states
necessarily have a nonzero degree of nonclassicality associated with
them \cite{footconj}.  This may be regarded as further support for
the physical significance of the decomposition into classical and nonclassical 
components.

Eq. (\ref{rex}) for pure states will now be proved as a special case of
the more general {\it inequality}
\begin{equation} \label{rhoeur}
\delta X\Delta P_{nc}\geq \hbar/2,
\end{equation}
holding for states described by density operators.  
While not an exact uncertainty relation, this inequality is
still much stronger than the corresponding Heisenberg inequality in Eq.
(\ref{heis}).  Not only is it saturated for {\it all} pure states (not
just the ``minimum uncertainty'' states), but it implies that 
properties such as Eq. (\ref{posit}) hold for {\it any} quantum state.

Inequality (\ref{rhoeur}) is an immediate consequence of 
Eq. (\ref{pdiff}) and the relations
\begin{equation}\label{rhoeq}
\frac{\hbar^2}{4(\delta X)^2} + \langle P_{cl}^2\rangle = \int dx\,
\frac{|\langle x|P\rho |x\rangle|^2}{\langle x|\rho |x\rangle}
\leq \langle P^2\rangle  , 
\end{equation}
which hold for all density operators $\rho$.  The equality in Eq.
(\ref{rhoeq}) is obtained by substituting Eqs. (\ref{pcl})
and (\ref{pclop}) for the classical momentum component $P_{cl}$,
and the representation
\begin{equation} \label{fisho}
(\delta X)^{-2} = -\frac{1}{\hbar^2}\int dx \, \frac{\langle
x|P\rho-\rho P|x\rangle^2}{\langle x|\rho |x\rangle} ,
\end{equation}
for the Fisher length, following from the definition of $\delta X$ in 
Eq. (\ref{fl}) and the
identity $(d/dx)\langle x|A|x\rangle=(i/\hbar)\langle x|[P,A]|x\rangle$
(derived by expanding in momentum eigenkets). 
The {\it in}equality in Eq. (\ref{rhoeq}) is obtained by defining the
states $|\mu\rangle =\rho^{1/2}P|x\rangle$,
$|\nu\rangle=\rho^{1/2}|x\rangle$, and using the Schwarz inequality 
\[
|\langle x|P\rho |x\rangle |^2 = |\langle\mu |\nu\rangle |^2 \leq
\langle\mu |\mu\rangle\langle\nu |\nu\rangle = \langle x|P\rho
P|x\rangle\langle x|\rho |x\rangle . \]
Remarkably, for the special case of a pure state, direct substitution of 
$\rho=|\psi\rangle\langle\psi |$ into the integral in Eq. (\ref{rhoeq})
yields equality on the righthand side, and hence the exact uncertainty 
relation Eq. (\ref{rex}).

Finally, note that a similar derivation may be given for the {\it
conjugate} uncertainty relation
\begin{equation}\label{xexact}
\Delta X_{nc} \delta P \geq \hbar/2,
\end{equation}
again saturated by pure states.  This relation 
similarly implies the Heisenberg
inequality; requires the variance in position to be infinite for
states with momentum distributions that are discontinuous or which
vanish over a continuous range of momentum values; and implies that
the variance of the nonclassical component of position is strictly
positive.

\subsection{Energy bounds}

Eqs. (\ref{pdiff}) and (\ref{rhoeur}) 
immediately yield the lower bound
\begin{equation}\label{enineq}
E\geq (8m)^{-1}\hbar^2(\delta X)^{-2} + \langle V\rangle
\end{equation}
for the average energy $E$ of any state. Moreover, 
from Eqs. (\ref{pclpsi}) and (\ref{rex}), this bound 
is saturated for all {\it real
wavefunctions}, such as energy eigenstates.  It follows that bounds
for energy may be obtained via corresponding bounds on the Fisher length
$\delta X$.

For example, consider the case of the one-dimensional Coulomb potential
$V(x)=-Zq^2/|x|$.  From Eqs. (1) and (9) of Ref. \cite{romera} one has
the bound $(\delta X)^{-2}\geq 4\langle |x|^{-1}\rangle^2$, and hence
from Eq. (\ref{enineq}) the lower bound
\[
E\geq (2m)^{-1}\hbar^2\langle |x|^{-1}\rangle^2-Zq^2\langle
|x|^{-1}\rangle \]
for energy.  Minimising with respect to $\langle |x|^{-1}\rangle$ then
yields the lower bound
\[
E_0\geq -Z^2q^4m/(2\hbar^2) \]
for the groundstate energy.  The righthand side is, fortuitously, the correct
groundstate energy, and this result may be generalized to the
three-dimensional case via the formalism in Sec. IV.E below.

A number of upper and lower bounds for the Fisher length are given by
Romera and Dehesa \cite{romera}, and by Dembo et al. \cite{dembo}, which
yield corresponding bounds on energy.  Eq. (34) of the latter reference
provides an interesting connection between groundstate energy estimation
and the entropy of the position observable.  In particular, the
``isoperimetric inequality'' \cite{dembo}
\[
\delta X\leq (2\pi e)^{-1/2} e^S , \]
where $S=-\int dx\,p(x)\ln p(x)$ is the position entropy, 
implies via Eq. (\ref{enineq}) the general
entropic lower bound
\begin{equation}\label{entbound}
E\geq (4m)^{-1}\pi e\hbar^2 e^{-2S} +\langle V\rangle .
\end{equation}

Eq. (\ref{entbound}) may be exploited to estimate groundstate energies
by maximising the position entropy for a given value of $\langle
V\rangle$.  Note this gives a lower bound on $E_0$, in contrast to the
usual upper bounds provided by variational methods.  For example, for a
harmonic oscillator with $V(x)=m\omega^2x^2/2$, the entropy is well known
to be maximised for a given value of $\langle x^2\rangle$ by a Gaussian
distribution.  Substituting such a distribution into Eq.
(\ref{entbound}) and minimising with respect to $\langle x^2\rangle$
then yields the estimate $E_0\geq \hbar\omega /2$, where 
the righthand side is in fact the correct groundstate energy 
(because the groundstate probability distribution is indeed 
Gaussian).

As a further example of Eq. (\ref{entbound}), consider a particle
bouncing in a uniform gravitational field, with $V(x)=mgx$ for $x\geq
0$.  For a fixed value $\langle x\rangle =\lambda$ one finds that the
entropy is maximised by the exponential distribution $p(x)= \lambda^{-1}
\exp (-x/\lambda)$ ($x\geq 0$), yielding the lower bound
\[
E\geq \pi\hbar^2(4me\lambda^2)^{-1} +mg\lambda .\]
Minimizing with respect to $\lambda$ then gives the
estimate
\[
E_0\geq (3/2)[\pi/(2e)]^{1/3}(mg^2\hbar^2)^{1/3}\approx 1.249\, 
(mg^2\hbar^2)^{1/3} , \]
which is comparable to the exact value of
$(mg^2\hbar^2/2)^{1/3}a_0 \approx 1.856\, (mg^2\hbar^2)^{1/3}$ obtained
by solving the Schr\"{o}dinger equation \cite{flugge}, where $a_0$
denotes the first Airy function zero.  

\subsection{Phase, angular momentum and photon number}

The exact uncertainty relations
\begin{eqnarray} \label{jexact}
\delta \Phi\Delta J_{nc} & = & \hbar/2 ,\\ \label{nexact}
\delta\Phi \Delta N_{nc} & = & 1/2,
\end{eqnarray}
for phase and angular momentum and for phase and photon number
respectively, may be proved exactly as per Eq. (\ref{rex}) above, and
are valid for all pure states.  For more general states described by
density operators the righthand sides become lower bounds.

It follows, for example, that the variance of angular momentum 
is infinite for states with phase distributions which are discontinuous or
vanish over some interval.  Similarly, the photon number variance is
infinite for states with a discontinuous phase distribution
\cite{footphaseint}.  Conversely, consider the case of a photon
number eigenstate.  From Eq. (\ref{nvar}) it follows that both the
classical and nonclassical fluctuations in photon number vanish, and
hence from the exact uncertainty relation in Eq. (\ref{nexact}) that 
the Fisher length $\delta\Phi$ is infinite, i.e., such states have a
uniform phase distribution (see Sec. III.B). Thus the exact uncertainty
relation in Eq. (\ref{nexact})
is sufficiently strong to exhibit the complementary nature of phase and
photon number.  Similar remarks may be of course be 
made for the case of angular
momentum.

The exact uncertainty relations may be used to derive the
Heisenberg-type inequalities \cite{hallphase} 
\begin{eqnarray}\label{jheis}
\Delta_\theta\Phi\Delta J & \geq & |1-2\pi p(\theta+\pi)|\hbar/2 , 
\\ \label{nheis}
\Delta_\theta\Phi\Delta N & \geq & |1-2\pi p(\theta+\pi)|/2 .
\end{eqnarray}
These follow directly from Eqs. (\ref{jexact}) and (\ref{nexact}),
using the modified Cramer-Rao inequality Eq. (\ref{phicramer}) and the
additivity of variances in Eqs. (\ref{jvar}) and (\ref{nvar}). 
Similar inequalities have been previously given by Pegg and Barnett
\cite{peggherm} and by Shapiro \cite{shap}. Note
that these inequalties are not of sufficient strength to draw
the conclusions obtained above from the {\it exact} uncertainty relations. 
Note further that for continuous phase distributions one can always
choose the reference angle $\theta$ such that the righthand sides
trivially vanish.

\subsection{Time and frequency}

In classical signal processing theory,
a signal is represented by a normalized time-varying
amplitude $a(t)$.  Since such signals typically obey linear propagation
laws, their analysis usually relies heavily on the frequency
representation $A(f)$ of $a(t)$, given by the Fourier transform
\begin{equation}
A(f) = \int dt\, a(t) e^{2\pi ift} .
\end{equation}
This relation is formally similar to the connection between
position and momentum amplitudes in quantum mechanics, and in particular
one has the well known time-frequency uncertainty relation
\begin{equation} \label{ft}
\Delta f \Delta t \geq (4\pi)^{-1}
\end{equation}
in analogy to the Heisenberg inequality Eq. (\ref{heis}).

The ``instantaneous frequency'' of the signal at time $t$ is defined as
\cite{sigpro}
\begin{equation} \label{finst}
f_{inst}(t) := (2\pi)^{-1} (d/dt)[\arg a(t)],
\end{equation}
which from Eq. (\ref{pcl}) is seen to be analogous to the classical
component of momentum.  Thus there is a corresponding decomposition of
frequency,
\begin{equation} \label{decomp}
f = f_{inst} + f_{fluc} ,
\end{equation}
into an instantaneous frequency component and a fluctuating frequency
component, analogous to Eq. (\ref{pdecomp}).  As per Sec. II.A, the
instantaneous frequency may be interpreted as the best possible estimate
of the frequency of the signal at a given time.

The purpose of this subsection is to point out the exact uncertainty
relation
\begin{equation} \label{fexact}
\Delta f_{fluc}\delta t = [{\rm Var}f -{\rm Var}f_{inst}]^{1/2}\delta t
= (4\pi)^{-1}
\end{equation}
for frequency and time.  This is formally equivalent to the relation for
position and momentum in Eq. (\ref{rex}), and may be proved in precisely
the same manner.  

The exact uncertainty relation  implies that the instantaneous frequency
$f_{inst}$ is a good estimate of frequency precisely when the ``Fisher
time" $\delta t$ is large.  Moreover, {\it causal} signals, defined to
be those for which $a(t)$ vanishes for all times less than some initial
time \cite{sigpro}, must have $\delta t=0$ (see Sec. III.A), and hence
it follows that $\Delta f=\infty$ for such signals.  The same conclusion
holds for any signal for which $a(t)$ is discontinuous or vanishes over
some interval.  Note that these conclusions cannot be derived from the weaker
inequality Eq. (\ref{ft}) (which itself follows as a consequence of the
exact uncertainty relation and the Cramer-Rao inequality in Eq.
(\ref{cramer})).

\subsection{Higher dimensions}

Exact uncertainty relations for vector observables are of interest not
only because the world is not one-dimensional, but because some physical
properties, such as entanglement, require more than one dimension for
their discussion.  It will therefore be indicated here how Eq.
(\ref{ex}) may be generalised to the case of $n$-vectors ${\bf X}$ and
${\bf P}$.  This case has also been briefly considered in Ref. \cite{hall}.
For simplicity only pure states will be considered.

First, one has the vector decomposition 
\begin{equation} \label{vecdecomp}
{\bf P} = {\bf P}_{cl} + {\bf P}_{nc}
\end{equation}
into classical and nonclassical components,
where ${\bf P}_{cl}$ commutes with ${\bf X}$, and
\begin{equation} \label{vecpcl}
{\bf P}_{cl}({\bf x}) = \langle {\bf x}|{\bf P}_{cl}|{\bf x}\rangle =
\frac{\hbar}{2i}\left( \frac{\nabla\psi}{\psi}-\frac{\nabla\psi^*
}{\psi^*}\right) = \hbar\nabla\left[ \arg\psi\right]
\end{equation}
is the best estimate of ${\bf P}$ from measurement value ${\bf X}={\bf
x}$ for state $\psi$ (one may also derive ${\bf P}_{cl}({\bf x})$ from
continuity equations or a Wigner function as per Secs. II.B and II.C).  
Note that since the vector components of ${\bf P}$ commute, as do the
vector components of ${\bf P}_{cl}$, then
\[
[P_{nc}^j,P_{nc}^k] = [P^j-P_{cl}^j,P^k-P_{cl}^k] =
(\hbar^2/i)(\partial_j\partial_k -\partial_k\partial_j)\left[ \arg\psi
\right]=0 ,\]
as claimed in Eq. (\ref{commcomp}). 
In analogy to Eqs. (\ref{pav}) and (\ref{pvar}) one may derive $\langle
{\bf P}\rangle = \langle{\bf P}_{cl}\rangle$ and the generalized linear
independence property
\begin{equation}\label{pcov}
{\rm Cov}({\bf P}) = {\rm Cov}({\bf P}_{cl}) + {\rm Cov}({\bf P}_{nc}) ,
\end{equation}
where the $n\times n$ covariance matrix of $n$-vector ${\bf A}$ is
defined by the matrix coefficients
\begin{equation}\label{cov}
\left[{\rm Cov}({\bf A})\right]_{jk} = \langle A_jA_k\rangle - \langle
A_j\rangle\langle A_k\rangle .
\end{equation}

Second, the notion of Fisher length for one dimension is generalized to
the matrix inverse
\begin{equation}
\label{fcov}
{\rm FCov}({\bf X}) := \left\{ \int d^nx\, p({\bf x})[\nabla \ln p({\bf
x})]\, [\nabla \ln p({\bf x})]^T\right\}^{-1} ,
\end{equation}
where ${\bf A}^T$ denotes the vector transpose of ${\bf A}$.  For the
case of one dimension this reduces to the square of the Fisher length
$\delta X$, just as the covariance matrix in Eq. (\ref{cov}) reduces to
the square of $\Delta A$.  Moreover, as per the covariance matrix, 
the matrix in Eq. (\ref{fcov}) is
real, symmetric and nonnegative. Finally, the
matrix is the inverse of the ``Fisher information'' matrix of statistical
estimation theory \cite{cox}.  For these reasons ${\rm FCov}({\bf X})$
will be referred to as the {\it Fisher covariance matrix} of ${\bf X}$.
One has the generalized Cramer-Rao inequality \cite{cox}
\begin{equation}\label{covcramer}
{\rm Cov}({\bf X}) \geq {\rm FCov}({\bf X}) ,
\end{equation}
with equality for Gaussian distributions.

One may show by direct calculation of ${\rm Cov}({\bf P}_{cl})$ that the
generalized exact uncertainty relation
\begin{equation}\label{covexact}
{\rm FCov}({\bf X})\, {\rm Cov}({\bf P}_{nc}) = (\hbar/2)^2I_n 
\end{equation}
holds for all pure states, where $I_n$ denotes the $n\times n$ unit
matrix.  The corresponding Heisenberg matrix inequality follows
immediately from Eqs. (\ref{pcov}), (\ref{covcramer}) and
(\ref{covexact}) as 
\begin{equation}\label{covheis}
{\rm Cov}({\bf X})\, {\rm Cov}({\bf P}) \geq (\hbar/2)^2I_n .
\end{equation}

The exact uncertainty relation, being a symmetric matrix equality,
comprises $n(n+1)/2$ independent equalities.  One may always choose $n$
of these equalities as corresponding to the diagonal elements of the
matrix equality obtained by multiplying Eq. (\ref{covexact}) on the left
by the inverse of ${\rm FCov}({\bf X})$. This yields a generalization
of the one-dimensional
exact uncertainty relation Eq. (\ref{ex}) for each {\it individual}
vector component of ${\bf X}$ and ${\bf P}$.  
A further 
choice is to take the square root of the determinant of both sides
of Eq. (\ref{covexact}), to give the corresponding ``volume'' equality
\begin{equation}\label{covvol}
\delta V_{\bf X}\Delta V_{{\bf P}_{nc}}=(\hbar/2)^n ,
\end{equation}
where the Fisher volume $\delta V$ and the covariance volume $\Delta V$ 
are defined as the square roots of the determinants of the respective
covariance matrices.  
For $n=1$ this relation reduces to
Eq. (\ref{ex}).

\section{ENTANGLEMENT AND CORRELATION}

Consider now the case of two one-dimensional particles, with respective
position and momentum observables $(X^{(1)},P^{(1)})$ and
$(X^{(2)},P^{(2)})$.  Such a system corresponds to $n=2$ in Sec. IV.E,
and the corresponding nonclassical momentum components associated with
wavefunction $\psi$ follow from Eqs. (\ref{vecdecomp}) and (\ref{vecpcl})
as
\begin{equation} \label{pentang}
P^{(1)}_{nc} = P^{(1)}-\hbar\frac{\partial \arg\psi(x_1,x_2)}{
\partial x_1},\hspace{1cm} 
P^{(2)}_{nc} = P^{(2)}-\hbar\frac{\partial \arg\psi(x_1,x_2)}{
\partial x_2} .
\end{equation}
For entangled states (e.g., a superposition of two product states), 
it follows that the
nonclassical momentum of particle 1 will typically depend on the position
observable of particle 2, and vice versa.  Hence if some
unitary transformation (e.g., a position displacement) is 
performed on the {\it second} particle, then the 
nonclassical momentum of the {\it first} particle is typically 
changed.

The decomposition into classical and nonclassical components is
therefore nonlocal:  the decomposition of a single-particle observable 
typically depends upon actions performed on another
particle with which the first is entangled.  Conversely, all such decompositions
are invariant under actions performed
on a second {\it un}entangled particle.  
The nonlocality inherent in quantum entanglement is thus 
reflected to some degree by the nonlocality
of classical/nonclassical decompositions.

The exact uncertainty relation corresponding to the decomposition of
momentum in Eq. (\ref{pentang}) is given by the matrix equality of Eq.
(\ref{covexact}), with $n=2$.  This leads to three independent
inequalities, as discussed in Sec. IV.E, two of which may be chosen as 
as generalizations of the exact uncertainty relation
in Eq. (\ref{ex}) for each individual particle.
The third independent inequality could, for example, be chosen as the
volume inequality in Eq. (\ref{covvol}).  However, 
a different choice provides an interesting connection with the Pearson
correlation coefficient of classical statistics. In particular, this
coefficient is defined for two compatible observables $A$ and $B$, 
in terms of the coefficients $C_{jk}$ of the corresponding
covariance matrix ${\rm Cov}(A,B)$, by \cite{cox}
\begin{equation} \label{pearson}
r_P(A,B) := C_{12}/(C_{11}C_{22})^{1/2} , 
\end{equation}
and provides a measure of the degree to which $A$ and $B$
are linearly correlated. It ranges between -1 (a high degree of
linear correlation with negative slope) and +1 (a
high degree of linear correlation with positive
slope).  One may analogously define the ``Fisher'' correlation
coefficient in terms of the coefficients $C^F_{jk}$ of the corresponding
Fisher covariance matrix
${\rm FCov}(A,B)$, with 
\begin{equation}\label{fisher}
r_F(A,B) := C^F_{12}/(C^F_{11}C^F_{22})^{1/2}  .
\end{equation}
This again provides a measure of correlation ranging between -1 and +1,
and is equal to the Pearson correlation coefficient for all Gaussian
distributions.

The third equality may now be
chosen as the simple correlation relation
\begin{equation} \label{corr}
r_P(P^{(1)}_{nc},P^{(2)}_{nc}) + r_F(X^{(1)},X^{(2)}) = 0 , 
\end{equation}
as may be verified by direct calculation from Eq. (\ref{covexact}).
Thus, for example, if the nonclassical momentum components of particles
1 and 2 are positively correlated then the position observables are
negatively correlated, and vice versa.  More generally, the degree of
nonclassical momentum correlation is seen to be
precisely determined by the degree
of position correlation.  The exact uncertainty relation in Eq.
(\ref{covexact}) thus constrains both uncertainty {\it and} 
correlation.

A nice example is provided by the approximate EPR
state
\[
\psi(x_1,x_2) = K e^{-(x_1-x_2-a)^2/4\sigma^2} 
e^{-(x_1+x_2)^2/4\tau^2} e^{ip_0(x_1+x_2)/(2\hbar)} ,\]
where $K$ is a normalisation constant and $\sigma <<1<<\tau$ in 
suitable units.  One may then
calculate
\begin{eqnarray*}
\langle X^{(1)}-X^{(2)}\rangle & = & a,\hspace{1cm}
{\rm Var}(X^{(1)}-X^{(2)}) = \sigma^2 << 1 ,\\
\langle P^{(1)}+P^{(2)}\rangle & = & p_0,\hspace{1cm}
{\rm Var}(P^{(1)}+P^{(2)}) = \hbar^2/\tau^2 <<1 ,
\end{eqnarray*}
and hence $\psi$ is an approximate eigenstate of the relative position
and the total momentum, i.e., one may write
\begin{equation}\label{approx}
X^{(1)}-X^{(2)}\approx a,\hspace{1cm} P^{(1)}+P^{(2)}\approx p_0 .
\end{equation}
This state is thus an approximate version of the (nonnormalizable) ket
considered by Einstein, Podolsky and Rosen in connection with
the completeness of the quantum theory \cite{epr}.
 
For state $\psi$ one finds from Eq. (\ref{vecpcl}) that the classical
components of momentum are constant, each being equal to $p_0/2$. 
Hence one has ${\rm Cov}{\bf P}_{nc} ={\rm Cov}{\bf P}$ from Eq.
(\ref{pcov}).  Then, since equality holds in Eq. (\ref{covcramer}) for
Gaussian distributions, the exact uncertainty relation corresponding to
$\psi$ follows from Eq. (\ref{covexact}) as
\begin{equation} \label{epruncert}
{\rm Cov}(X) {\rm Cov}(P) = (\hbar/2)^2 I_n  .
\end{equation}
Eq. (\ref{corr}) reduces to  
(recalling that $r_P$ and $r_F$ are equivalent for
Gaussian distributions) the correlation relation
\[
r_P({\bf X}) + r_P({\bf P}) = 0 .\]
This latter result 
is consistent with Eq. (\ref{approx}), which implies that 
$X^{(1)}$ and $X^{(2)}$ are highly positively correlated for state
$\psi$ [$r_P({\bf X})\approx 1$], while $P^{(1)}$ and $P^{(2)}$ are
highly negatively correlated [$r_P({\bf P})\approx -1$].

Finally, it is of interest to consider the effect of measurements on the
approximate EPR state $\psi$.  First, for a position
measurement on particle 2, with result $X^{(2)}=x$, the state of
particle 1 collapses to the wavefunction obtained 
by substituting $x_2=x$ and renormalising.  It follows that the the
classical momentum component $P_{cl}^{(1)}$ remains equal to $p_0/2$.
Hence the momentum decomposition of particle 1 is not altered by
knowledge of $X^{(2)}$.

Conversely, for a momentum measurement on particle 2 with result
$P^{(2)}=p$, one finds via straightforward calculation of the
appropriate Gaussian integrals that the state of particle 1 collapses
to the wavefunction
\[
\psi(x_1|P^{(2)}=p) =
K'e^{-(x_1+a/2)^2/(\sigma^2+\tau^2)/4}e^{i\tilde{p}x_1/\hbar} ,\]
where $K'$ is a normalisation constant and 
\[
\tilde{p} = \frac{\sigma^2p+\tau^2(p_0-p)}{\sigma^2+\tau^2} .\]
It follows that the classical momentum component $P^{(1)}_{cl}$ 
is {\it not} invariant
under a measurement of $P^{(2)}$, changing from $p_0/2$ to
$\tilde{p}$.  Hence there is a ``nonlocal'' effect on the
classical/nonclassical decomposition of momentum for particle 1, brought
about by a measurement of $P^{(2)}$.  This effect is a
reflection of the strong correlation between $P^{(1)}$ and $P^{(2)}$ for
state $\psi$.  In particular, note that since $\sigma <<1<<\tau$, one has
$\tilde{p}\approx p_0-p$, as might well be expected from Eq.
(\ref{approx}).

\section{NON-CONJUGATE AND DISCRETE \\OBSERVABLES}

Exact uncertainty relations can be formally extended in a very
general way to arbitrary pairs of Hermitian observables.  Unfortunately,
the physical significance of such an extension is not entirely
clear, as will be seen below.  However, for the case of a complete set of
mutually complementary observables on a finite Hilbert space it will be
shown that results in the literature provide a very satisfactory
form of exact uncertainty relation.

First, consider the case of {\it any} two observables $A$ and $B$
represented by Hermitian operators, and for state $\rho$ define
\begin{equation}\label{bcl}
B^A_{cl} := \sum_a |a\rangle\langle a|\frac{\langle a|B\rho +\rho
B|a\rangle/2}{\langle a|\rho |a\rangle} .
\end{equation}
Here $|a\rangle$ denotes the eigenket of $A$ with eigenvalue $a$, and
the summation is replaced by integration for continuous ranges of
eigenvalues.

Clearly the above expression generalises Eqs. (\ref{pcl}) and
(\ref{pclop}), and indeed $B^A_{cl}$ may be interpreted as providing the
best estimate of $B$ compatible with measurement of $A$ on state $\rho$.
Note that $A^A_{cl}=A$, i.e., $A$ is its own best estimate.
One may further define $B^A_{nc}$ via the decomposition
\[ B = B^A_{cl} + B^A_{nc}, \]
and obtain the relations
\[
\langle B\rangle = \langle B^A_{cl}\rangle, \hspace{1cm} {\rm Var}B =
{\rm Var}B^A_{cl}+{\rm Var}B^A_{nc} \]
for state $\rho$, in analogy to Eqs. (\ref{pav}) and (\ref{pvar}).

If one is then prepared to define the quantity $\delta_B A$ by
\[
(\delta_B A)^{-2} = \sum_a \frac{\langle a|(i/\hbar)[B,\rho]|a\rangle^2
}{\langle a|\rho |a\rangle} , \]
in analogy to Eq. (\ref{fisho}), then precisely as per the derivation of
Eq. (\ref{rhoeur}) one may show that
\begin{equation} \label{aexact}
(\delta_B A)\,\Delta B^A_{nc} \geq \hbar/2,
\end{equation}
with equality for all pure states.

Thus there is a very straightforward generalisation of Eq. (\ref{ex}) to
arbitrary pairs of observables.  A difficulty is, however, to provide
a meaningful statistical interpretation of $\delta_BA$.  Note in
particular that, unlike the Fisher length $\delta X$, this quantity is
not a functional of the probability distribution $\langle a|\rho
|a\rangle$ in general.  Possibly, noting the commutator which appears in
the definition of $\delta_BA$, one can interpret this quantity as a
measure of the degree to which a measurement of $A$ can distinguish
between $B$-generated translations of state
$\rho$, i.e., between unitary transformations of the form $e^{ixB/\hbar}\rho
e^{-ixB/\hbar}$ \cite{helhol}.  Here such an attempt will not be made.

Finally, it is pointed out that a rather different type of
exact uncertainty relation exists for a set of $n+1$ mutually
complementary observables $A_1, A_2, \dots ,A_{n+1}$ on an
$n$-dimensional Hilbert space.  Such sets are defined by the property
that the distribution of any member is uniform for an eigenstate of any
other member, and are known to exist when $n$ is a power of a prime
number \cite{wootters}.  
As an example one may choose $n=2$, and take $A_1$, $A_2$
and $A_3$ to be the Pauli spin matrices.

Let $L$ denote the collision length
of probability distribution $\{p_1, p_2, \dots p_n\}$, defined by \cite{hell}
\[
L := 1/ \sum_j (p_j)^2 . \]
Note that $L$ is equal to 1 for a distribution concentrated on a single
outcome, and is equal to $n$ for a distribution spread uniformly over
all $n$ possible outcomes. It hence provides a direct measure of the spread
of the distribution over the space of outcomes \cite{hell}.

One may show that \cite{ivan}
\begin{equation} \label{mut}
\sum_i 1/L_i = 1+{\rm tr}[\rho^2] \leq 2 ,
\end{equation}
where $L_i$ denotes the collision length of observable $A_i$ for state
$\rho$.  
This reduces to a strict equality for all pure states, and 
thus provides an exact uncertainty relation for the
collision lengths of any set of $n+1$ mutually complementary
observables.  For example, if $L_j=1$ for some
observable $A_j$ (minimal uncertainty), then $L_i=n$
for all $i\neq j$ (maximal uncertainty).
Ivanovic has shown that Eq. (\ref{mut}) can be
used to derive an entropic uncertainty relation for the $A_i$
\cite{ivan}, while Brukner and Zeilinger have interpreted Eq.
(\ref{mut}) as an additivity property of a particular 
``information'' measure \cite{zeil}.

\section{CONCLUSIONS} 

It has been shown that the uncertainty principle has in fact an element
of certainty:  the lack of knowledge about an observable is, for any
wavefunction, {\it precisely} determined by the lack of knowledge about
the conjugate observable.  The measures of lack of knowledge must of
course be chosen appropriately (as the nonclassical fluctuation strength
and the Fisher length).  What is remarkable is that such measures
can be chosen at all.

The exact uncertainty relations in Eqs. (\ref{ex}), (\ref{jexact}),
(\ref{nexact}) and (\ref{covexact}) are formal consequences of the
Fourier transformations which connect the representations of conjugate
quantum observables.  Hence they may be extended to any domain in which
such transformations have physical significance.  This includes, for
example, the time-frequency domain considered in Sec. IV.D, as well as
Fourier optics and image processing.

It would be of interest to determine whether exact uncertainty relations
exist for relativistic systems.  One is hampered in direct attempts 
by difficulties associated with one-particle interpretations of the
Klein-Gordon and Dirac equations.  It would perhaps therefore be more
fruitful to first consider extensions to general field theories.

Finally, note that the definition of the Fisher covariance matrix in Eq.
(\ref{fcov}) suggests an analogous definition of a ``Wigner'' covariance
matrix ${\rm WCov}$, defined via the coefficients of its matrix inverse
\[ [{\rm WCov}^{-1}]_{jk} := \int d^{2n}z\, W^{-1} \frac{\partial
W}{\partial z_j} \frac{\partial W}{\partial z_k} .\]
Here $W$ denotes the Wigner function of the state, and ${\bf z}$ denotes
the phase space vector $({\bf x},{\bf p})$.  It would be of interest to
determine to what degree this matrix is well-defined, and to what extent 
its properties characterise nonclassical features of quantum
states. 

{\bf Acknowledgment}
I thank Marcel Reginatto for constant encouragement and many helpful comments on
the subject matter of this paper.
%\newpage


\begin{thebibliography}{99}
\bibitem[19]{footphase} Strictly speaking this is a second moment of the
distribution, and the variance should, in analogy to the nonperiodic
case, be identified with the minimum value of ${\rm Var}_\theta \Phi$ as
$\theta$ is varied over a range of $2\pi$ \cite{hallphase}.
\bibitem[21]{footconj} It was conjectured in Ref. \cite{hall} that the
{\it joint} nonclassicality $\Delta X_{nc}\Delta P_{nc}$ is not only
strictly positive, but is bounded below by
$\hbar/2$ for the case of pure states.  This conjecture is, however, violated by the wavefunction
$K\exp[-x^2+i\alpha x^2]$, where $K$ and $\alpha$ are real constants.
\bibitem[25]{footphaseint} The semibounded spectrum of photon number
implies that the conjugate phase distribution cannot vanish over any
interval (phase is fuzzy!) \cite{shap}.
\end{thebibliography}
\end{document}